# COMPUTATIONAL ALGORITHM FOR ORBIT AND MASS DETERMINATION OF VISUAL BINARIES

MOHAMED SHARAF [1], MOHAMAD NOUH [2], ABDEL NABY SAAD [2],
MAGDY ELKHATEEB [2], SOMAYA SAAD [2]

[1] *Department of Astronomy, Faculty of Science*
*King Abdul-Aziz University, Jeddah, Saudi Arabia*

[2] *Department of Astronomy*
*National Research Institute of Astronomy and Geophysics, Helwan, Egypt*
*E-mail: abdo_nouh@hotmail.com*

*Abstract*. In this paper we introduce an algorithm for determining the orbital elements and individual masses of visual binaries. The algorithm uses an optimal point, which minimizes a specific function describing the average length between the least-squares solution and the exact solution. The objective function to be minimized is exact, without any approximation. The algorithm is applied to Kowalsky's method for orbital parameter computation, and to Reed's method for the determination of the dynamical parallax and individual masses. The procedure is applied to A 1145 and ADS 15182.

## 1. INTRODUCTION

Visual binaries are almost the only objects in the universe whose masses can be determined directly. To determine masses, the semimajor axis and the period of the visual orbit must be known. Given the modern computers, many authors proposed automated methods of orbit determination. Docobo (1985) and Docobo et al. (1992) used Cid's method, which implies 3½ observation points, in the same manner as Thiele-Innes-van den Bos method. All observations are used simultaneously only in the final stage.

Many proposed techniques or methods require a sufficiently precise set of orbital parameters as a first approximation. So far, Eichhorn (1985) and Eichhorn and Xu (1990) developed a method that needs an accurate initial orbit and all observations are used simultaneously in the improvement phase only.

Catovic and Olevic (1992) suggested an approach that consists of adding one falsified observation, which should be chosen so that the solution of the applied least squares is an ellipse. Consequently, by changing the position, one can draw a large number of elliptic orbits and the final decision of choosing the best one is left to the computer.

Pourbaix (1994) and Pourbaix and Lampens (1997) developed a method based on a function that quantifies the distance between observed and computed position. The simulated annealing method was successfully used to minimize it; in this way the best orbit is that one which minimizes the adopted function.

In this paper we develop an algorithm for orbit and mass determination of visual binaries. The algorithm uses an optimal point $(\rho_a, \theta_a)$, which minimizes a specific function and describes the average length between the least square solution and the exact one. The objective function to be minimized is exact, without any approximation. We used Kowalsky's method (Smart 1953) for the determination of the orbital parameters, and Reed's (1984) method for the determination of the dynamical parallax and individual masses. The algorithm is applied to two visual systems A 1145 and ADS 15182.

## 2. BASIC EQUATIONS

The condition equation of the apparent ellipse can be written as

$$Ax^2 + 2Hxy + By^2 + 2Fx + 2Gy + 1 = 0, \qquad (2.1)$$

which involves five $(A, B, G, H, F)$ of the seven necessary orbital elements as fitting parameters (e.g., Smart 1953). In Equation (2.1), $x$ and $y$ are related to the observed positions $(\rho, \theta)$ via

$$x = \rho \cos \theta, \qquad (2.2)$$

$$y = \rho \sin \theta, \qquad (2.2)$$

where $\theta$ is the position angle (in degrees), whereas $\rho$ stands for the angular separation (in arcseconds), or

$$x = AX + FY,$$
$$y = BX + GY,$$

while $X$ and $Y$ are given as:

$$X = \cos E - e,$$
$$Y = \sqrt{1 - e^2} \sin E,$$

where $e$ and $E$ denote the eccentricity and the eccentric anomaly, respectively.

When applying equation (2.1), one must take into account both the inaccuracy of the input data and the fact that the observed arcs are usually small.

Least-squares method seems to be one of the most powerful techniques for solving the condition equations, but at the same time is exceedingly critical. This is because the least-squares estimate suffers from the deficiency of the mathematical optimization techniques that give point estimates; the estimation procedure has not been built-in detecting and controlling techniques for the sensitivity of the solution to the optimization criterion of $\sigma^2$ to be minimum. Due to these two problems, we cannot guarantee that the observed arc could give an elliptic solution. And if the ellipticity is not questioned, we can expect more than one orbits which would be sufficiently accurate. Thus, and according to the above mentioned features, *we shall try to find a method, which controls the best ellipse fitted to equation (2.1)*.

### 2.1. ANALYTICAL ERROR CRITERION

If $C$ and $C'$ denote the exact and the least-squares solution of equation (2.1), respectively, there may exist a situation in which there are many significantly different vectors $C'$ that reduce the variance $\sigma^2$ to an acceptable small value. It seems that the averaged squared distance between $C$ and $C'$ can be used as a criterion for the acceptable solution of equation (2.1).

Let the normal equations (e.g., Kopal and Sharaf 1980) be defined by

$$GC' = b. \qquad (2.4)$$

Then, the accuracy of the solution is controlled by the condition of the system (2.4) (of being well- or ill-conditioned). More precisely, the essential criterion is the size of the minimum eigenvalue ($\lambda_{min}$) of $G$.

Following Kopal and Sharaf (1980), the value which minimizes the difference between $C$ and $C'$ is adopted as

$$Q(\rho, \theta) = \sigma^2 \sum_{i=1}^{5} \frac{1}{\lambda_i}, \qquad (2.5)$$

or

$$Q(\rho, \theta) = \sigma^2 S. \qquad (2.6)$$

The analytic expression of $S$ can be easily obtained. Let

$$\det| G - \lambda I | = \prod_{i=1}^{5} (\lambda - \lambda_i) = \alpha_0 + \alpha_1 \lambda + \alpha_2 \lambda^2 + \alpha_3 \lambda^3 + \alpha_4 \lambda^4 + \lambda^5, \quad (2.7)$$

be the characteristic polynomial of the matrix $G$. Then

$$S = \left(\prod_{j=1}^{5} \lambda_j\right)^{-1} \sum_{j=1}^{5} \prod_{i=1, i\neq j}^{5} \lambda_i = -\frac{\alpha_1}{\alpha_0}, \quad (2.8)$$

where

$$\alpha_0 = \lambda_1\lambda_2\lambda_3\lambda_5 + \lambda_1\lambda_2\lambda_4\lambda_5 + \lambda_1\lambda_3\lambda_4\lambda_5 + \lambda_2\lambda_3\lambda_4\lambda_5 + \lambda_1\lambda_2\lambda_3\lambda_4,$$
$$\alpha_1 = -\lambda_1\lambda_2\lambda_3\lambda_4\lambda_5.$$

The elements of $G$ are given by

$$g_{ij} = g_{ji} = \sum_{k=1}^{5} \phi_{ik}\phi_{jk}, \quad (2.9)$$

where

$$\phi_{1k} = \sum_{k=1}^{N} x_k^2, \; \phi_{2k} = \sum_{k=1}^{N} 2x_k y_k, \; \phi_{3k} = \sum_{k=1}^{N} y_k^2, \; \phi_{4k} = \sum_{k=1}^{N} 2x_k, \; \phi_{4k} = \sum_{k=1}^{N} 2y_k,$$

so that $\alpha_0$ and $\alpha_1$ are obtained as a combination of the coefficients $g_{ij}$. Then, the best solution of equation (2.1) is encountered at the minimum value of the function $Q(\rho, \theta)$. According to the foregoing definition of $Q$, it is clear that it depends on all observations.

## 2.2. MINIMIZATION OF $Q$

The global minimum of the objective function $Q$ is carried out using Simulated Annealing Method (Metropolis 1953). This method has already been successfully applied to the determination of the orbital parameters of visual binaries (Pourbaix 1994, 1998; Pourbaix and Lampens, 1997). The procedure is based mainly on the minimization of a function that features the difference between observed and calculated positions.

In the original method, the objective function we want to minimize is $\varepsilon$ (the energy of any physical system). If $\varepsilon_i$ is the energy in the configuration $i$ of the system and $\varepsilon_{i+1}$ that in the configuration $i+1$, and if $\varepsilon_{i+1} \leq \varepsilon_i$, the new configuration is obviously accepted; but, it is also accepted if

$$r \leq -(\varepsilon_{i+1} - \varepsilon_i)/(K \cdot Temp),$$

where $r$ is a random number in the range 0 to 1, $K$ is the Boltzmann constant, and *Temp* is the temperature of the system. To use this method we need a value analogous to the temperature (*Temp*) and a way to reduce it. A value of *Temp* between 0.5 and 1 was deduced; it was reduced by 0.999 at each step to visit a larger space of the function range. We used the algorithm described by Press et al. (1992) which resorted to a sophisticated procedure for the random point generator based on the Modified Simplex Method (Nedler and Mead 1965). If the algorithm is stopped as soon as the local minimum is reached, we restart using a new simplex.

As stated by Pourbaix (1998), one cannot expect to obtain the global minimum, so a local search algorithm must be used to tune the minimum. To this end we used Powell's method (Press et al. 1992), which needs $n$ conjugate directions in the $n$-dimensional space ($n = 2$ in our two-variable case: $\rho$, $\theta$). The base vectors are a good simple choice for these directions.

### 2.3. MASS DETERMINATION

The dynamical parallax and individual masses $(M_A, M_B)$ of the components of binary systems, in solar units, can be computed (Reed 1984) from the apparent magnitudes $(m_a, m_b)$, the orbital period $P$ (years), and the semimajor axis of the true orbit $a''$ (arcseconds), in seconds of arc, via the equations

$$\log M_B = [m_b - \alpha - \tfrac{5}{3}\log(1 + \Delta/\beta) - \lambda]/(\tfrac{5}{3} + \beta), \qquad (2.10)$$

and

$$M_A = M_B(10^\alpha), \qquad (2.11)$$

where $\Delta = m_a - m_b$, $\lambda = (10/3)\log P - 5\log a'' - 5$, $\alpha = 4.6$, $\beta = -9.5$.

Then the distance in parsecs can be determined from

$$r = \frac{(M_A + M_B)^{1/3} P^{2/3}}{a''}. \qquad (2.12)$$

### 3. COMPUTATIONAL ALGORITHM

**Purpose:** To calculate the orbital elements $a$, $e$, $i$, $\Omega$, $\omega$, $P$, $T$ of the visual binary by Kowalsky's method.

**Input:** $t_i$, $\theta_i$, $\rho_i$, $i = \overline{1, N}$.

**Output:** $a$, $e$, $i$, $\Omega$, $\omega$, $P$, $T$, $Q$, $\sigma^2$.

**Computational Sequence:**
1. Compute, for $i = \overline{1, N}$, $x = \rho \cos \theta$, $y = \rho \sin \theta$.
2. Solve equation (2.1) by least-squares, which yield $A$, $B$, $F$, $G$, $H$, $Q_1$ (calculated numerically), $\sigma^2$.
3. Compute $a$, $e$, $i$, $\Omega$, $\omega$, $P$, $T$ using Kowalsky's method.
4. Determination of the initial orbit is completed.
5. For the list of observations do:
   - Minimize the objective function $Q$ (equation (2.5)); this step yields $Q, (\rho_a, \theta_a)$.
   - Add $(\rho_a, \theta_a)$ to the list of observations.
6. Go to step 1 to compute the final orbit ($a$, $e$, $i$, $\Omega$, $\omega$, $P$, $T$, $Q$).
7. Compute the individual masses and parallax from equations (2.10), (2.11), (2.12).

The algorithm is completed.

**Remark.** The value of $Q_1$ (computed at step 2) and $Q$ (computed at step 5) must be the same for the final orbit.

### 4. APPLICATION

We elaborated a program in FORTRAN 77 for the computations needed by the above procedures together with Kowalsky's method for the determination of the orbital elements. It was applied to the binary systems A 1145 and ADS 15182.

#### 4.1. A 1145

Two orbits for the system A 1145 had been computed by da Silva and Balca (1968) and Heintz (1979). The result of Heintz includes moments after periastron. Table 1 lists the orbital elements, the standard deviation $\sigma^2$, the minimum $Q$ value and the additional point $(\rho_a, \theta_a)$. For the initial orbit, $\sigma^2$ and $Q$ were computed numerically from the least-square solution of the condition equation, while for the final orbit we used the optimized values of the analytic function (equation (2.5)). Also, for sake of comparison, we computed the function $D$ adopted by Pourbaix (1994) for the two orbits. This function describes the difference between the observed and calculated positions as

$$D = \frac{\sum_i\{[(\rho_o - \rho_c)/\mu_\rho]^2 + [(\theta_o - \theta_c)/\mu_\theta]^2\}}{\sum_i w_i},$$

where, for visual observations, $\mu_\rho = \mu_\theta = 0.5$, $w = 1$.

As it is clear from the result, the present orbit, controlled by $Q$, also reduces the function $D$ to its minimum value. Dynamical parallax and masses are also confirming our orbital parameters.

*Table 1*

**Orbital elements, dynamical parallax and masses of A 1145**

| Parameter | Present orbit | Heintz orbit |
|---|---|---|
| $a''$ | 0.499 | 0.415 |
| $e$ | 0.28 | 0.27 |
| $i$ (°) | 135.3 | 130.2 |
| $\Omega$ (°) | 9.7 | 22.3 |
| $\omega$ (°) | 272.8 | 297.5 |
| $P$ (years) | 142.6 | 137 |
| $T$ | 1962.31 | 1967 |
| $\sigma^2$ | 0.0057 | ... |
| $Q$ | 0.2379 | ... |
| $D$ | 5.29E-2 | 4.544 |
| $\rho_a$ (") | 0.443 | ... |
| $\theta_a$ (°) | 351.6 | ... |
| $a^3/P^2$ | 4.458E-6 | 3.808E-6 |
| $\pi$ (") | 1.093E-2 | 1.025E-2 |
| $M_A$ (solar masses) | 2.072 | 2.143 |
| $M_B$ (solar masses) | 1.339 | 1.385 |
| $M_{AB}$ (solar masses) | 3.411 | 3.528 |

4.2. ADS 15182

The first orbit for this system has been computed by Baize (1980). Due to the great discrepancies between the observed and the calculated positions, Jasinta (1996) computed a provisional orbit for the system using an epoch of observation up to 1993. Table 2 is the corresponding table, which listed our results and has the same designation as Table 1.

*Table 2*

**Orbital elements, dynamical parallax and masses of ADS 15182**

| Parameter | Present orbit | Jasinta orbit |
|---|---|---|
| $a''$ | 0.236 | 0.21 |
| $e$ | 0.335 | 0.58 |
| $i$ (°) | 140.27 | 149.8 |
| $\Omega$ (°) | 150.516 | 72.8 |
| $\omega$ (°) | 330.62 | 262.6 |
| $P$ (years) | 149.62 | 133.81 |
| $T$ | 2003.5 | 1998.5 |
| $\sigma^2$ | 0.00332 | ... |
| $Q$ | 0.6289 | ... |
| $D$ | 1.243 | 10.339 |
| $\rho_a$ (") | 0.278 | ... |
| $\theta_a$ (°) | 128.76 | ... |
| $a^3/P^2$ | 5.85E-7 | 5.17E-7 |
| $\pi$ (") | 5.9E-3 | 5.62E-3 |
| $M_A$ (solar masses) | 1.453 | 1.492 |
| $M_B$ (solar masses) | 1.384 | 1.421 |
| $M_{AB}$ (solar masses) | 2.837 | 2.913 |

## 5. CONCLUSION

We developed a computational algorithm for determining the orbital parameters of visual binaries. This was done by adding an optimal point $(\rho_a, \theta_a)$ which minimizes a specific function describing the average length between the least-squares solution and the exact solution. It should be mentioned that the objective function to be minimized is exact, without any approximation. Moreover, it has a geometrical meaning: the average square distance. On this basis, individual masses and dynamical parallaxes have also been determined. The numerical results prove the applicability of the method in obtaining accurate elements depending on the minimum value of $Q$, without resorting to the usual criterion of minimum $\sigma^2$. Dynamical parallax and the individual masses are also included and agree well with the values determined by other authors.